\documentstyle[12pt]{article}
\begin{document}
\centerline
{{\large {\bf Graded Symmetry Algebras of Time-Dependent Evolution}}} 
\centerline
{{\large {\bf 
Equations and Application to the Modified KP equations}}} 
\vskip 2mm

\centerline {Wen-Xiu Ma\footnote{Email: 
wenxiuma@uni-paderborn.de}, 
R. K. Bullough\footnote{Email: Robin.Bullough@umist.ac.uk}
and P. J. Caudrey\footnote{Email: Philip.Caudrey@umist.ac.uk}  
}  
\centerline{Department of Mathematics, 
University of Manchester} 
\centerline{Institute of Science and Technology, PO Box 88,
Manchester M60
1QD, UK}   

\vskip 5mm

\centerline{{\it Dedicated to Prof. W. I. Fushchych on the occasion of his
 60th birthday}}

\newtheorem{thm}{Theorem}
\newtheorem{lemma}{Lemma}
\newcommand{\R}{\mbox{\rm I \hspace{-0.9em} R}}        
\def\be{\begin{equation}}
\def\ee{\end{equation}}
\def\ba{\begin{array}}
\def\ea{\end{array}}
\def\bea{\begin{eqnarray}}
\def\eea{\end{eqnarray}}
\def\la {\lambda}
\def \part {\partial}
\def \al {\alpha}
\def \de {\delta}

\renewcommand{\theequation}{\thesection.\arabic{equation}}

\setlength{\baselineskip}{18pt}

\begin{abstract}
By starting from known graded Lie algebras, including Virasoro 
algebras, new kinds of time-dependent evolution equations are found 
possessing graded symmetry algebras. The modified KP equations are taken 
as an illustrative example: new modified KP equations with $m$ arbitrary 
time-dependent coefficients are obtained possessing symmetries involving $m$ 
arbitrary functions of time. A particular graded symmetry algebra for the 
modified KP equations is derived in this connection homomorphic to the 
Virasoro algebras.  
\end{abstract}

\section{Introduction}
\setcounter{equation}{0}

Symmetries are one of the important and currently active areas in soliton 
theory. They are closely connected with the integrability of corresponding 
nonlinear equations. For a given evolution equation 
\be u_t=K(t,x,u) \ \ (u=u(t,x), u_t\equiv \frac{\partial u}{\partial t}),
\label{ee} \ee 
a vector field $\sigma (t,x,u)$ is called its symmetry (or generalized 
symmetry) if
$\sigma (t,x,u)$ satisfies its linearized equation 
\be \frac {d \sigma (t,x,u)}{dt}
=K'[\sigma ]\ {\rm or} \ \frac {\part \sigma(t,x,u)}
{\part t}=[K,\sigma ],\label{sd}\ee
where the prime means the Gateaux derivative:
\be K'[\sigma ]=\left.\frac {\part }{\part \varepsilon }
\right |_{\varepsilon=0}
K(u+\varepsilon \sigma )\ee 
and the Lie product $[\cdot,\cdot]$ is defined by 
\be [K,\sigma ]=K'[\sigma ]-\sigma '[K]=
\left.\frac {\part }{\part \varepsilon }
\right |_{\varepsilon=0}
K(u+\varepsilon \sigma )-\left.\frac {\part }{\part \varepsilon }
\right |_{\varepsilon=0}
\sigma (u+\varepsilon K )
.\label{Liep}\ee
Actually, the symmetries defined above are infinitesimal generators of
one-parameter groups of invariant transformations of $u_t=K(t,x,u)$. If 
a vector field
 $\sigma $ does not depend on the time variable $t$, the condition
of $\sigma $ being a symmetry of (\ref{ee}) becomes a 
very simple equality, namely $[K,\sigma ]=0,$ which means that a symmetry $\sigma$
only needs to commute with 
the vector field $K$ generating the considered equation. However if 
the $\sigma $ depend on $t$, then the problem is not so simple.
Some specific methods for dealing with this case were introduced, for example,
in Refs. \cite{OevelF} \cite{Fuchssteiner1983} \cite{Fokas}
\cite{Ma1991}.

Note that the above definition of symmetry may also be viewed as
\be [u_t-K(t,x,u), \sigma (t,x,u)]=0,\label{nsd}\ee
 where the Lie product should be understood as 
the one in the extended vector field Lie algebra including the time variable
$t$, not just including the space variable $x$, as in \cite{FuchssteinerO}.
A Lie homomorphism $\textrm{exp}(\textrm{ad}_{T})$ of the vector field Lie
algebra with a suitable vector field 
$T=T(t,x,u)$ may be applied to the discussion of the integrability 
of time-dependent evolution equations.
Here $\textrm{ad}_{T}$ denotes the adjoint map
of the vector field $T$ and thus we have \[(\textrm{ad}_{T})S=[T,S]\ 
\ \textrm{for any vector field}\  
S=S(t,x,u).\]
Fuchssteiner observed \cite{Fuchssteiner1993} 
that if the Lie homomorphism $\textrm{exp}(\textrm{ad}_{T})$
acts on (\ref{nsd}), a new and significant result may be reached which states 
that a new evolution equation
\be u_t=\textrm{exp}(\textrm{ad}_{T})K+\sum_{i=0}^\infty 
\frac {(\textrm{ad}_{T})^i}
{(i+1)!}\part _tT\ee
has a symmetry $\textrm{exp}(\textrm{ad}_{T})\sigma (t,x,u) $, 
when $\sigma (t,x,u)$ is 
a symmetry of 
$u_t=K(t,x,u)$. This is because we have 
\bea  &&\textrm{exp}(\textrm{ad}_{T}) 
[u_t-K(t,x,u), \sigma (t,x,u)]\nonumber \\
&=& [\textrm{exp}(\textrm{ad}_{T})
(u_t-K(t,x,u)),\textrm{exp}(\textrm{ad}_{T})\sigma (t,x,u)] \nonumber\eea
and 
\[ \textrm{exp}(\textrm{ad}_{T})
(u_t)= u_t-\sum_{i=0}^\infty \frac {(\textrm{ad}_{T})^i}
{(i+1)!}\part _tT.\] 
Here we require, of course, that the relevant series converge.

The present paper aims at the construction of time-dependent 
evolution equations which possess graded symmetry algebras
and most particularly centreless Virasoro algebras. The basic 
tools we will adopt in this paper are 
the above observation by Fuchssteiner \cite{Fuchssteiner1993} and the 
Lax operator algebra method in \cite{Ma1992a}.
The result of the analysis gives rise to various
 concrete realizations of graded Lie algebras.
As an illustrative example,  
two graded Lie algebras are presented for the modified KP equations. 
Moreover an application of our result for constructing evolution equations 
with arbitrary time varying coefficients and a graded symmetry algebra 
involving these arbitrary coefficients is given for the modified KP equations. 
Some concluding remarks are given in the last section.

\section{Variable-coefficient equations from Virasoro algebras}
\setcounter{equation}{0}

We take the centreless Virasoro algebra:
\be \left \{ \ba {l} 
\left [K_{l_1},K_{l_2}\right]=0,
\vspace{2mm} \\
\left[K_{l_1},\rho_{l_2}\right]=(l_1+\gamma )K_{l_1+l_2},
\vspace{2mm} \\
\left [\rho_{l_1},\rho_{l_2}\right]=(l_1-l_2)\rho_{l_1+l_2}, \ea
\right. \label{Va} 
\ee
in which the vector fields $K_l,\, \rho_l$ do not depend explicitly on
the time variable $t$ and the $\gamma $ is a fixed constant.
Note that here the space variable $x$ may belong to $\R^p$ or $Z^p$ and
$u(x,t)$ may belong to $\R ^q$ generally. 
If we define an operator $\Phi $ as 
\be \Phi K_l=K_{l+1},\ \Phi \sigma _l=\sigma _{l+1}, \ee
in which $\sigma _l $ is a symmetry of $u_t=K_l $, then this operator
 $\Phi $ is hereditary over the above Virasoro algebra, i.e. 
it is to satisfy the equality
\be \Phi ^2[K, S] +[\Phi K,\Phi S]-\Phi \{ [\Phi K, S]+[K,\Phi S]\}=0\ee
for any vector fields $K,S$ belonging to that Virasoro algebra 
(see \cite{Fuchssteiner1996} for example).

We first consider an equation 
\be u_t=\al _1(t)K_{i_1}\label{K1}\ee
with $\al _1(t)$ an arbitrary function of time.
Choose a key vector field as
\be T_1=\beta _1(t)K_{i_1}\ \textrm{where}\ \frac {\part }{\part t}
\beta _1(t)=\al _1(t).\ee 
Then we have 
\[ \textrm{exp}(\textrm{ad}_{T_1})(u_t)=u_t+[T_1,u_t]=u_t-\al _1(t)K_{i_1} \]
so that an application of $\textrm{exp}(\textrm{ad}_{T_1})$
to the zero equation $u_t=0$ yields the equation (\ref{K1}) considered.
Since any vector field which does not depend explicitly on $t$ 
is a symmetry of $u_t=0$, we have, in particular,
the symmetries $K_l,\ \rho_l$.
Further by applying $\textrm{exp}(\textrm{ad}_{T_1})$ to these symmetries,
we obtain two hierarchies of symmetries of (\ref{K1})
\bea && \textrm{exp}(\textrm{ad}_{T_1}) K_l=K_l, \label{k1sh1}
\\ &&\sigma _{l,i_1}=
\textrm{exp}(\textrm{ad}_{T_1})\rho_l=\rho_l+[T_1,\rho_l]\nonumber \\&& =
\beta_1(t)[K_{i_1},\rho_l]+\rho_l=\beta_1(t)(i_1+\gamma )K_{i_1+l}+\rho_l,
\label{k1sh2}
\eea
which also constitute the same  Virasoro algebra as (\ref{Va}).
Of course, we might generate other symmetries of (\ref{K1}) from 
any vector field $\rho=\rho (x,u)$ which causes the series $\textrm{exp}
(\textrm{ad}_{T_1})\rho $
to converge. Here we give only the two sorts of such symmetries,
because any other symmetries, just based on Virasoro algebras, are not at 
all clear as symmetries. 

We next consider the more general equation 
\be u_t=\al _1(t)K_{i_1}+\al _2(t)K_{i_2}\label{K2}\ee
with two arbitrary functions of time $\al _1(t),\, \al _2(t)$.
We choose a key vector field 
\be T_2=\beta _2(t)K_{i_2} \ \textrm{where}\ \frac {\part }{\part t}
\beta _2(t)=\al _2(t).\ee 
Now we find that an application of $\textrm{exp}(\textrm{ad}_{T_2})$
to the evolution equation (\ref{K1}) 
and its symmetries $K_l,\,\sigma _{l,i_1}$
 yields 
the considered equation (\ref{K2}), together with its following symmetries:- 
\bea && \textrm{exp}(\textrm{ad}_{T_2}) K_l=K_l, \label{k2sh1}
\\ &&\sigma _{l,i_1i_2}=\textrm{exp}(\textrm{ad}_{T_2})\sigma _{l,i_1}
\nonumber\\ &&
=\textrm{exp}(\textrm{ad}_{T_2})[\beta_2(t)(i_2+\gamma )K_{i_2+l}]
+\beta_2(t)(i_2+\gamma )K_{i_2+l}\rho_l\nonumber \\ &&
=\beta_1(t)(i_1+\gamma )K_{i_1+l}+\beta_2(t)(i_2+\gamma )K_{i_2+l}+\rho _l,
\label{k2sh2}
\eea
which still constitute the same Virasoro algebra as (\ref{Va}).

In general, we can obtain 
the variable-coefficient evolution equation 
\be u_t=\al _1(t)K_{i_1}+\al _2(t)K_{i_2}+\cdots +\al _m(t)K_{i_m}
\label{Km}\ee
with $m$ given arbitrary time-dependent functions $\al _j(t),\, 1\le j\le m$,
and its 
two hierarchies of symmetries 
\bea && \textrm{exp}(\textrm{ad}_{T_m})\cdots
\textrm{exp}(\textrm{ad}_{T_1}) K_l=K_l, \label{kmsh1}
\\ &&\sigma _{l,i_1\cdots i_m}=\textrm{exp}(\textrm{ad}_{T_m})\sigma _{l,i_1
\cdots i_{m-1}}=\cdots \nonumber \\ &&
= \textrm{exp}(\textrm{ad}_{T_m})\cdots 
\textrm{exp}(\textrm{ad}_{T_1})\rho_l \nonumber \\
&&= \sum_{j=0}^m 
\beta_j(t)(i_j+\gamma )K_{i_j+l}+\rho _l,
\label{kmsh2}
\eea
where $T_j=\beta _j(t)K_{i_j},\ \frac {\part }{\part t}\beta _j(t)=\al _j(t),\ 
1\le j\le m $. 
The symmetries so obtained constitute 
a Virasoro algebra with the same commutation relations as (\ref{Va}):
\be \left \{ \ba {l} 
\left [K_{l_1},K_{l_2}\right]=0,
\vspace{2mm} \\
\left[K_{l_1},\sigma _{l_2,i_1\cdots i_m}
\right]=(l_1+\gamma )K_{l_1+l_2},
\vspace{2mm} \\
\left [\sigma _{l_1,i_1\cdots i_m},\sigma _{l_2,i_1\cdots i_m}\right]
=(l_1-l_2)\sigma _{l_1+l_2,i_1\cdots i_m}. \ea
\right. \label{sVa} 
\ee
This may also be directly checked. 
For example, we can calculate that
\bea && \left [\sigma_{l_1,i_1\cdots i_m},\sigma_{l_2,i_1\cdots i_m}\right]
 \nonumber \\ &  = &
\Bigl[\sum_{j=0}^m\beta_j(t)(i_j+\gamma )K_{i_j+l_1}+\rho _{l_1},
\sum_{j=0}^m\beta_j(t)(i_j+\gamma )K_{i_j+l_2}+\rho_{l_2}\Bigr]
\nonumber\\
&=&    
\Bigl[\sum_{j=0}^m\beta_j(t)(i_j+\gamma )
K_{i_j+l_1},\rho_{l_2}\Bigr]
+ \Bigl[\rho_{l_1},
\sum_{j=0}^m\beta_j(t)(i_j+\gamma )K_{i_j+l_2}\Bigr]
+\left[\rho_{l_1},\rho_{l_2}\right]
\nonumber \\
&=&(l_1-l_2) \sum_{j=0}^m\beta_j(t)(i_j+\gamma )K_{i_j+l_1+l_2}
+(l_1-l_2)\rho_{l_1+l_2}\nonumber\\
&=&  (l_1-l_2)\sigma_{l_1+l_2,i_1\cdots i_m}. \nonumber
\eea
Note that the symmetries $K_l$ are generally time-independent,
while at the same time, the symmetries $\sigma _{l,i_1\cdots i_m}$ include
$m$ given arbitrary functions of time and so are 
 time-dependent in the same way as our considered equation (\ref{Km}).
The symmetries $\sigma _{l,i_1\cdots i_m}$ contain the generators of
Galilean invariance and 
invariance under scale transformations \cite{Fushchych} \cite{FushchychT}.
On the other hand, the symmetry algebra (\ref{sVa}) provides a new explicit 
realization of the original Virasoro algebra ({\ref{Va}).
We can still take the Virasoro algebra (\ref{sVa}) to be a starting algebra.
However any new result is really no more than that
we have already reached.

If we choose $\al _i(t), 1\le i\le m$, to be polynomials in time,
then the evolution equation (\ref{Km}) and its
 symmetries (\ref{kmsh2}) are of polynomial-in-time type (see \cite{MaBCF}). 
Therefore we may see that there exist 
higher-degree polynomial-in-time dependent symmetries for many 
evolution equations in $1+1$ dimensions. This 
is itself an interesting result in the symmetry theory of evolution equations,
because a soliton equation in $1+1$ dimensions usually has only 
master symmetries of the first order 
(the reader is referred to \cite{Fuchssteiner1983} for a definition of master symmetry). 
Furthermore, our derivation
does not refer to any particular choices of dimensions and space variables.
Hence the evolution equation (\ref{Km}) may be not only both
continuous ($x\in \R ^p$) and discrete ($x\in Z^p$), but also both $1+1$ 
($p=1$) and higher dimensional ($p>1$).

It is well known that there are many 
integrable equations which possess a centreless Virasoro 
algebra (\ref{Va}) (see \cite{ChenL} \cite{MaF} \cite{ChengLB} 
\cite{Ma1990} \cite{OevelFZ}
\cite{MaZ} for example). 
Among the most famous examples are the KdV hierarchy
in the continuous case and the Toda lattice hierarchy
in the discrete case.
According to the result above, we can say that
a KdV-type equation 
\be u_t=t^{n_1}K_0+t^{n_2}K_1=t^{n_1}u_x+t^{n_2}(u_{xxx}+6uu_x)\ \ 
(n_1,n_2\in Z /\{-1\} )\ee
possesses a hierarchy of time-dependent symmetries
\be  \sigma _{l,01}
=\frac  {t^{n_1+1} }{2(n_1+1)}
K_{l}+\frac {3 t^{n_2+1} }{2(n_2+1)}K_{l+1}+\rho_l.
\ee
Here the vector fields $K_l,\,\sigma _l,$ are defined by
\[ K_l=\Phi^l u_x,\ \rho_{l}=\Phi ^l(u+\frac 12 xu_x), \ l\ge 0,\]
in which the $\Phi $ is a well known hereditary 
operator
\[ \Phi = \part _x^2+ 4u+2u_x\part _x^{-1}.\]
They constitute a centreless Virasoro algebra (\ref{Va}) with $\gamma =
\frac 12 $ \cite{Ma1990} \cite{Ma1992} and thus so do the symmetries 
$K_l,\, \sigma_{l,01}$. The symmetries $\sigma_{l,01}$ are
of the polynomial-in-time type 
when $n_1\ge0 $ and $n_2\ge 0 $, and they
are of the Laurent polynomial-in-time type
when $n_1\le -2 $ or $n_2\le -2 $. The latter 
 is worthy of notice, since a 
time-independent evolution equation does not have such 
symmetries \cite{Ma1991}.

We can also conclude that a Toda-type lattice equation 
\bea && (u(n))_t=\left(\ba {c} p(n) \vspace{2mm} \\ v(n)   \ea  \right) _t
=K_0+t^{n_1}K_1+t^{n_2}K_0\ \ (n_1,n_2\in Z/\{-1\} )\nonumber\\
&& =(1+t^{n_2})\left(\ba {c} v(n)-v(n-1) \vspace{2mm}\\     
v(n)(p(n)-p(n-1))\ea  \right)\nonumber \\
&& \quad  + t^{n_1}
\left(\ba {c} p(n)(v(n)-v(n-1))+v(n)(p(n+1)-p(n-1)) \vspace{2mm}\\     
v(n)(v(n-1)-v(n+1))+v(n)(p(n)^2-p(n-1)^2)\ea  \right)
\eea
possesses a hierarchy of time-dependent symmetries
\be 
\sigma _{l,010}=tK_l+\frac{t^{n_1+1}}{n_1+1}K_{l+1}+\frac { t^{n_2+1}}{n_2+1}
 K_l+\rho _l.\ee
Here the vector fields $K_l,\ \rho_l$ are defined by
\[
K_l=\Phi ^{l}K_0,\ \rho_{l}=\Phi ^l \rho_0,\
K_0=\left ( \ba {c} v-v^{(1)}\vspace{1mm}
\\ v(p-p^{(-1)})\ea \right ),\ \rho_0=\left ( \ba {c} p\vspace{2mm}
\\ 2v\ea \right ),  \ l\ge 0,  \]
 in which  the hereditary operator  $\Phi$ is given by
\[ \Phi=\left ( \ba {cc} p&(v^{(1)}E^2-v)(E-1)^{-1}v^{-1}\vspace{1mm}
\\  v(E^{-1}+1)& v(pE-p^{(-1)})(E-1)^{-1}v^{-1}\ea \right ).
\]
Here we have used a normal shift operator $E$: $(Eu)(n)=u(n+1)$ and 
$u^{(m)}=E^mu, m \in Z$.
These discrete 
vector fields $K_l$ 
together with the discrete vector fields $\rho _l$
constitute a centreless Virasoro algebra (\ref{Va})
with $\gamma =1$ \cite{MaF} and thus the symmetry 
Lie algebra consisting of $K_l$ and $\sigma _{l,010}$
has the same commutation relations as that
Virasoro algebra.

\section{Variable-coefficient equations from general graded algebras}
\setcounter{equation}{0}

In this section, we consider more general 
algebraic structures by starting 
from a general graded Lie algebra.
In keeping with the notation in \cite{Ma1992a},
let us write a graded Lie algebra consisting 
of vector fields not depending explicitly on the time variable $t$ 
as follows:
\be E(R)=\sum_{i=0}^\infty E(R_i),\ [E(R_i), E(R_j)]\sqsubseteq E(R_{i+j-1}),
\ i,j\ge 0,\label{gradeda}\ee 
where $E(R_{-1})=0$ and the Lie product $[\cdot,\cdot]$ is defined by
(\ref{Liep}). Note that such a graded Lie algebra is called a master
Lie algebra in \cite{Ma1992a} since it is actually similar to a semi-graded 
Lie algebra under the group $<Z,*>$ with $i*j=i+j-1$, but not 
a graded Lie algebra as defined in \cite{Kac}. 
It itself looks like a $W_{1+\infty}$ algebra and 
includes a Virasoro algebra $E(R_0)+E(R_1)$ and a $W_\infty$ type
algebra $\sum_{i=1} ^\infty E(R_i)$ as subalgebras. The $W_\infty$ and 
$W_{1+\infty}$ type algebras broadly appear in conformal field theory
and in 2-dimensional quantum gravity \cite{BulloughC1} \cite{BulloughC2}. 
However here we focus on applications to symmetries of variable-coefficient 
evolution equations.

Consider a 
variable-coefficient evolution equation
\be u_t=\al _1(t)K_1+\al _2(t)K_2+\cdots
+\al _m(t)K_m,\label{gKm}
\ee 
with $m$ given arbitrary time-dependent functions $\al _j(t),\ 1\le j\le m$.  
As in section 2 we can first 
choose a key vector field to be
\be T_1=\beta _1(t)K_{1}\ \textrm{where}\ \frac {\part }{\part t}
\beta _1(t)=\al _1(t).\ee 
We then observe that the application of $\textrm{exp}(\textrm{ad}_{T_1})$
to the zero equation $u_t=0$ and its symmetries
$\rho_l, \rho_l\in E(R_l)$ yields the evolution
equation $u_t=\al _1(t)K_{i_1}$
and its  symmetries
\be \sigma _{1}(\rho_l)=
\textrm{exp}(\textrm{ad}_{T_1})\rho_l=
\sum_{j=0}^l
\frac {\beta _1^j}{j!}(\textrm{ad}_{K_{1}})^j\rho_l.
\label{gk1sh}
\ee
Note that the series $\textrm{exp}(\textrm{ad}_{T_1})\rho_l$ is truncated 
at the $l+1$-th term.
These symmetries also
constitute the same graded Lie algebra as (\ref{gradeda}).
We next choose a vector field 
\be T_2=\beta _2(t)K_{2} \ \textrm{where}\ \frac {\part }{\part t}
\beta _2(t)=\al _2(t).\ee 
and make an application of $\textrm{exp}(\textrm{ad}_{T_2})$ to 
$u_t=\al _1(t)K_{1}$ and its symmetries $\sigma _{1}(\rho_l)$.
This way we obtain the following evolution equation
\[ u_t=\al _1(t)K_{1}+\al _2(t)K_{2}\] 
and its symmetries 
\be
\sigma _{12}(\rho_l)=\textrm{exp}(\textrm{ad}_{T_2})
\sigma _{1}(\rho_l)=\sum_{0\le j_1+j_2\le l}
\frac {\beta _1^{j_1}\beta _2^{j_2}}{j_1!j_2!}(\textrm{ad}_{K_{1}})^{j_1}
(\textrm{ad}_{K_{2}})^{j_2}\rho_l.
\label{gk2sh}
\ee
Note that here we interchanged the position of $(\textrm{ad}_{K_{1}})^{j_1}$
and $(\textrm{ad}_{K_{2}})^{j_2}$, because we have 
\[ [K_{1},[K_{2},\rho_l]]=[K_{2},[K_{1},\rho_l]].\]
In general, we can obtain
the variable-coefficient evolution equation (\ref{gKm})
and its following  symmetries 
\bea &&
\sigma _{1\cdots m}(\rho_l)=\textrm{exp}(\textrm{ad}_{T_m})
\sigma _{1\cdots {m-1}}(\rho_l)
=\cdots = \textrm{exp}(\textrm{ad}_{T_m})\cdots 
\textrm{exp}(\textrm{ad}_{T_1})\rho_l \nonumber \\
&&= \sum_{0\le j_i+\cdots +j_m\le l}\frac{ \beta _1^{j_1}\cdots 
 \beta _m^{j_m}}{j_1!\cdots j_m!}(\textrm{ad}_{K_{1}})^{j_1}\cdots
(\textrm{ad}_{K_{m}})^{j_m}\rho _l,
\label{gkms}
\eea
where $\frac {\part }{\part t}\beta _j(t)=\al _j(t),\, 1\le j\le m.$
These symmetries still constitute 
a graded Lie algebra as at (\ref{gradeda}), that is, we have
\be [\sigma _{1\cdots m}(\rho_{l_1}), 
\sigma _{1\cdots m}(\rho_{l_2})]=\sigma _{1\cdots m}
([\rho_{l_1},\rho_{l_2}]),\ \rho_{l_1}\in 
E(R_{l_1}),\ \rho_{l_2}\in E(R_{l_2}).  \label{liepp}
\ee
Therefore 
the map 
\be \sigma  _{1\cdots m}: E(R)\to \sigma_{1\cdots m}(E(R)),\   
\rho_l\mapsto \sigma _{1\cdots m}(\rho_l),\, \rho_l\in E(R_l),\ee
 is a Lie homomorphism
between the graded Lie algebra (\ref{gradeda}) and the graded symmetry algebra
\be \sigma  _{1\cdots m}(E(R))=\sum_{j=0}^\infty
\sigma  _{1\cdots m}(E(R_j)).
\ee
Of course, we may also directly prove 
the Lie homomorphism property (\ref{liepp}).
We use mathematical induction to prove the required result. 
The proof is the following:
\bea 
&& [\sigma _{1\cdots m}(\rho _{l_1}), \sigma _{1\cdots m}(\rho _{l_2})]
\nonumber \\ &=&
\Bigl [ \sum_{0\le j_1+\cdots +j_m\le l_1} \frac {\beta _1^{j_1}\cdots 
\beta _m^{j_m}}{j_1!\cdots j_m!}(\textrm{ad}_{K_1})^{j_1}\cdots 
(\textrm{ad}_{K_m})^{j_m}\rho_{l_1},\nonumber \\ &&
\qquad 
 \sum_{0\le j'_1+\cdots +j'_m\le l_2} \frac {\beta _1^{j'_1}\cdots 
\beta _m^{j'_m}}{j'_1!\cdots j'_m!}(\textrm{ad}_{K_1})^{j'_1}\cdots 
(\textrm{ad}_{K_m})^{j'_m}\rho_{l_2}
\Bigr ]
\nonumber \\ &=&
\Bigl [ \sum_{j_m=0}^{l_1}\frac {\beta _m^{j_m}}{j_m!}
(\textrm{ad}_{K_m})^{j_m}\sigma_{1\cdots m-1}(\rho_{l_1}),
\sum_{j'_{m}}^{l_2}\frac {\beta _m^{j'_m}}{j'_m!}
(\textrm{ad}_{K_m})^{j'_m}\sigma_{1\cdots m-1}(\rho_{l_2})
\Bigr ]
\nonumber \\ &=&
 \sum_{j_m=0}^{l_1}\sum_{j'_{m}=0}^{l_2}\frac{\beta _m^{j_m+j'_m}}{j_m!j'_m!}
\Bigl [(\textrm{ad}_{K_m})^{j_m}\sigma_{1\cdots m-1}(\rho _{l_1}),
(\textrm{ad}_{K_m})^{j'_m}\sigma_{1\cdots m-1}(\rho _{l_2})
\Bigr ]
\nonumber \\ &=&
\sum_{j''_m=0}^{l_1+l_2-1}\frac {\beta _m^{j''_m}}{j''_m!}\sum_{j_m+j'_m=j''_m}
\frac {{j''_m!}}{j_m!j'_m!}\Bigl [
(\textrm{ad}_{K_m})^{j_m}\sigma_{1\cdots m-1}(\rho _{l_1}),
(\textrm{ad}_{K_m})^{j'_m}\sigma_{1\cdots m-1}(\rho _{l_2})
\Bigr ]
\nonumber \\ &=&
\sum_{j''_m=0}^{l_1+l_2-1}\frac {\beta _m^{j''_m}}{j''_m!}
(\textrm{ad}_{K_m})^{j''_m}\Bigl [\sigma_{1\cdots m-1}(\rho _{l_1}),
\sigma_{1\cdots m-1}(\rho _{l_2})
\Bigr ]
\nonumber \\ &=&
\sum_{j''_m=0}^{l_1+l_2-1}\frac {\beta _m^{j''_m}}{j''_m!}
(\textrm{ad}_{K_m})^{j''_m}\sigma_{1\cdots m-1}([\rho _{l_1},\rho _{l_2}])
\nonumber \\ &=&\sigma_{1\cdots m}([\rho _{l_1},\rho _{l_2}]).
\nonumber 
\eea 
Here in the last but one step, we have used the induction assumption.
Generally the symmetries $\sigma _{1\cdots m}(\rho _l)$ are 
time-independent when $l=0$ and 
 time-dependent when $l\ge 1$.

A graded Lie algebra has been exhibited for the time-independent 
KP hierarchy \cite{OevelF}
in \cite{Fuchssteiner1983}
\cite{Ma1992a}, which includes a centreless Virasoro algebra 
\cite{ChengLB} \cite{ChenLL}. 
Therefore we may also generate the corresponding 
graded Lie algebra of  time-dependent symmetries
for a resulting new set of variable-coefficient KP equations, and this 
is done in our paper \cite{MaBCF}.
In the next section, we shall go on to construct another graded Lie algebra,
which is related to the {\it modified} KP hierarchy. 

\section{Application to the modified KP equations}
\setcounter{equation}{0}

We first obtain a graded symmetry Lie algebra for the modified KP equations
and then give an application of the theory presented in the last section
to symmetries of the modified KP equations.
Let us consider 
the $2+1$ dimensional spectral operator $L$ corresponding to the modified KP
hierarchy:
\be L= \part^2_x+u\part _x+\part_y,\ u=u(t,x,y), \ t,x,y\in \R. \ee 
Evidently, its Gateaux derivative operator reads as $L'[X]=X\part _x,$
 and thus is injective, i.e. if $L'[X_1]=L'[X_2],$ then $X_1=X_2$.

Choose the following
polynomial differential operators in $\part_x$
\be A=\sum _{k=1}^ma_k\part_x^k,\ a_k=a_k(x,y,u),\
m\ge1 \label{Aform}\ee
as candidates for Lax operators \cite{Ma1992a}.
Then we may make the following calculation
\bea 
 AL &=& \sum_{k=1}^ma_k\part _x^k (\part _x^2+u\part _x+\part _y)\nonumber \\
&=&\sum_{k=1}^ma_k\part _x^{k+2}+ \sum_{k=1}^ma_k\sum_{i=0}^k
{k \choose i} (\part _x^{k-i}u)\part _x^{i+1}+\sum_{k=1}^ma_k\part _x^k
\part _y \nonumber \\
& =&\sum_{k=1}^ma_k\part _x^{k+2}+\sum_{k=1}^ma_k(\part _x^ku)\part _x+
\sum_{k=1}^ma_k\sum_{i=1}^k{k \choose i}
 (\part _x^{k-i}u)\part _x^{i+1}+\sum_{k=1}^ma_k\part _x^k
\part _y \nonumber \\
&   =&\sum_{k=1}^ma_k\part _x^{k+2}+\sum_{k=1}^ma_k(\part _x^ku)\part _x+
\sum_{i=1}^m\sum_{k=i}^m{k \choose i}a_k
 (\part _x^{k-i}u)\part _x^{i+1}+\sum_{k=1}^ma_k\part _x^k
\part _y \nonumber \\
   & =&\sum_{k=1}^ma_k\part _x^{k+2}+\sum_{k=1}^ma_k(\part _x^ku)\part _x+
\sum_{k=1}^m\sum_{i=k}^m{i \choose k}a_i
 (\part _x^{i-k}u)\part _x^{k+1}+\sum_{k=1}^ma_k\part _x^k
\part _y,
\nonumber \\
LA&  =& (\part _x^2+u\part _x+\part _y)\sum_{k=1}^ma_k\part _x^k
\nonumber \\  
&= &\sum_{k=1}^m(a_{kxx}\part _x^k+2a_{kx}\part _x^{k+1}+a_k\part _x^{k+2})
\nonumber \\ 
&& + \sum_{k=1}^m(ua_{kx}\part _x^k+ua_k\part _x^{k+1})+
\sum_{k=1}^m(a_{ky}\part _x^k+a_k\part _x^k\part _y).
\nonumber 
\eea
Here the coefficient ${k \choose i}$ denotes the standard binomial 
coefficient, i.e. ${k \choose i}=\frac {k!}{i!(k-i)!}$. Further we have 
\bea  [A,L]&=&AL-LA  
\nonumber \\
& = &\sum_{k=1}^m a_{k}(\part _x^ku)\part _x+\sum_{k=1}^m\sum_{i=k}^m
{i \choose k}a_i(\part _x^{i-k}u)\part _x^{k+1}\nonumber \\
&& - \sum_{k=1}^m(a_{kxx} +ua_{kx}+a_{ky})\part _x^k-\sum_{k=1}^m(
2a_{kx}+ua_k)\part _x^{k+1}\nonumber \\
& =&\sum_{k=1}^ma_k(\part _x^ku)\part _x+\sum_{k=1}^{m-1}\sum_{i=k+1}^m
{i \choose k}a_i(\part _x^{i-k}u)\part _x^{k+1}
\nonumber \\
&& -\sum_{k=1}^m(a_{kxx} +ua_{kx}+a_{ky})\part _x^k-\sum_{k=1}^m2a_{kx}\part 
_x^{k+1}. 
\eea 
Now we can see that the differential operator $A$
of the form (\ref{Aform}) is a Lax operator, i.e. there exists a vector
field $X$ such that $[A,L]=L'[X]=X\part _x $ if and only if the $a_k,\
1\le k\le m,$ satisfy the following equations 
\be \left \{\ba {l}  a_{mx}=0,\vspace{2mm}\\
{\displaystyle \sum_{i=k}^m{i \choose k-1} a_i(\part _x^{i-k+1}u)-
(a_{kxx} +ua_{kx}+a_{ky}+2a_{k-1,x})=0,}\vspace{2mm}\\
\qquad \qquad \qquad\qquad \qquad \qquad \qquad\qquad\quad k=m,m-1, \cdots, 2;
\ea \right. \label{keye}\ee
and the vector field $X$ must be the following,
\be X=\sum_{k=1}^ma_k(\part _x^ku)-(a_{1xx}+ua_{1x}+a_{1y}).\label{X}
\ee
This vector field is called an eigenvector field corresponding to the 
Lax operator $A$ of (\ref{Aform}). 

We denote by $W$ the following space: 
\be W=\bigl \{ f+g|\, f=\sum_{i,j\ge 0} c_{ij} x^iy^j,\ c_{ij}\in C,\
g=\sum  _{i,j,l\ge 0,\, k\in Z}
c_{ijkl}x^iy^j\part_x^k\part_y^lu,\ c_{ijkl}\in C
\bigr \},\ee
in which $\part^{-1}_x=\frac12 \bigl ( \int
_{-\infty}^x-\int_x^\infty\bigr) \, dx'$,
and we introduce the inverse operator $\part_x^{-1}$ of $\part_x$ over
the space $W$ as follows, namely 
\be \part_x^{-1} h = \part^{-1}_x (f+g) = \int _0 ^x f\, dx' +\frac12
\bigl( \int _{-\infty}^x-\int_{x}^\infty \bigr) g\, dx',\  h=f+g\in
W.\ee
Furthermore suppose that $C[y]$ denotes all polynomials in $y$, and
$C_n[y]\ (n\ge0),$ all polynomials in $y$ with degrees no greater than
$n$. Define a group of coefficients as follows 
\be a_k=a_{k1}+a_{k2},\ a_{k1}=a_k|_{u=0},\ a_{k2}=a_k-a_{k1},\ 1\le
k\le m,\ee
where the $a_{ki}$ are given recursively (from $m$ to $1$) by
\be \left \{ \ba {l} a_{m1}=c_m,\ a_{m2}=0;\vspace{2mm}\\
a_{k-1,1}=-\frac12 ( \part _x+\part _x ^{-1} \part_y )a_{k1}+c_{k-1},
\vspace{2mm}\\
{\displaystyle a_{k-1,2}=-\frac12 ( \part _x+\part _x ^{-1} \part_y )
a_{k2}-\frac12 \part _x^{-1}(ua_{kx})+\frac12 \part _x^{-1}
\sum ^m_{i=k} {i \choose k-1}a_i
(\part _x^{i-k+1} u),}\vspace{2mm}\\
\qquad\qquad\qquad\qquad\qquad \qquad\qquad\qquad \qquad\qquad\qquad \quad 
  k=m,m-1, \cdots 2,\ea \right. \label{a_krecursionrelation}\ee
where  $c_k\in C[y],\ 1\le k\le m.$ Obviously we see that
$a_k=a_{k1}+a_{k2}\in W,\ 1\le k\le m.$ For every set of $c_k\in
C[y],\ 1\le k\le m,$ 
we can uniquely determine a
Lax operator $A=\sum_{k=1}^m a_k\part_x^k=\sum_{k=1}^m (a_{k1}+a_{k2})
\part_x^k$ through 
(\ref{a_krecursionrelation}),
which operator is also written as 
$A=P(c_1,\cdots,c_m)$ in order to show the set of $c_k$.
We denote by $R$ and $R_i$ the following spaces of Lax operators
\bea  && R =\bigl\{ A=
P(c_1,\cdots,c_m)|\, m\ge1,\ c_k\in C[y],\ 1\le k\le m\bigr\}, \label{R}\\&&
R_i =\bigl
\{ A= P(c_1,\cdots,c_m)|\, m\ge1,\ c_k\in C_i[y],\ 1\le k\le m\bigr\},
i\ge 0.\label{R_i}\eea

In what follows, we want to prove that $R=\sum_{i=0}^\infty R_i$ is a
graded Lie algebra, namely to prove 
under the operation 
\be \lbrack\!\lbrack A, B \rbrack\!\rbrack =A'[Y]-B'[X]+[A,B],\label{pd}\ee
where $[A,L]=L'[X],\ [B,L]=L'[Y],$
that we have 
\be \lbrack\!\lbrack R_i, R_j \rbrack\!\rbrack \sqsubseteq R_{i+j-1}, R_{-1}=0,
\ i,j\ge 0.\label{R_iR_j}\ee 
We recall that we already have a product property \cite{Ma1992a}
\be [\lbrack\!\lbrack A, B \rbrack\!\rbrack,L]=L'[\,[X,Y]\,],
\label{Laxoperatorpp}\ee 
which will be used to derive a graded symmetry algebra later on. 
First by (\ref{a_krecursionrelation}), we immediately
 obtain the following two basic results.
\begin{lemma} \label{lemma1} Let $
 A= P(c_1,\cdots,c_m)
=\sum_{k=1}^m a_k\part_x^k
\in R ,$ and set 
\be  a_{k3 }=a_{k1}-c_k=a_k|_{u=0}-c_k,\ 1\le k\le m.\label{a_{k3}}\ee
 Then we have (1) $A$ is multi-linear with respect to $c_1,\cdots, c_m$;
(2) $A|_{u=0}=\sum_{k=1}^m(c_k+a_{k3})\part _x^k;$ (3) if $
A|_{u=0}=0,
$ then $A=0$.\end{lemma}

\begin{lemma} \label{lemma2} Let $
 A= P(c_1,\cdots,c_m)
=\sum_{k=1}^ma_k\part_x^k\in R_i$ and the $a_{k3}$ be defined by (\ref{a_{k3}}).
Then when $i=0$, $a_{k3}=0, 1\le k\le m$;
and when $i\ge1$, $  a_{m3}=0 $ and 
$ a_{k3},\ 1\le k\le m-1,$ are polynomials in $x,y$ with degrees
less than $i$ with respect to $y$.
\end{lemma}

In order  to verify that $R =\sum _{i=0}^\infty R_i$ is a graded Lie
algebra under the operation ({\ref{pd}), we go on to
 derive two other results.

\begin{lemma} \label{lemma3}
If $A\in R_0$, then $[A,L]|_{u=0}=X|_{u=0}\part _x=0;$ and if
$A\in R_i\, (i\ge1),$ then the coefficient of the differential operator
$[A,L]|_{u=0}$ is a polynomial in $x,y$ with degrees less than $i$
with respect to $y$.
\end{lemma}
\noindent {\bf Proof:} 
Assume that 
$$A= P(c_1,\cdots,c_m)
=\sum_{k=1}^ma_k\part_x^k.$$ 
By noting (\ref{X}), we discover that
\bea &&
[A,L]|_{u=0}=X|_{u=0}\part _x=-( a_{1xx}+ua_{1x}+a_{1y})|_{u=0}\part _x
\nonumber\\&&
=-(a_{11xx}+a_{11y})\part _x=-[a_{13xx}+(a_{13}+c_1)_y]\part _x,\nonumber \eea
where $ a_{13}$ is defined by (\ref{a_{k3}}).
Therefore the required result follows from
Lemma \ref{lemma2}, which completes the proof.
$\vrule width 1mm height 3mm depth 0mm$

\begin{lemma} \label{lemma4}
If $A,B\in R_0$, then $[A,B]|_{u=0}=0;$ and if
$A\in R_i,\,B\in R_j\  (i,j\ge0,\ i+j \ge1),$
then the coefficients of the differential operator
$[A,B] |_{u=0}$ are polynomials in $x,y$ with degrees less than $i+j$
with respect to $y$.
\end{lemma}
\noindent  {\bf Proof:} 
Suppose that
 $$A= P(c_1,\cdots,c_m)
=\sum_{k=1}^ma_k\part_x^k, B= P(d_1,\cdots,d_n)=\sum_{l=1}^nb_l\part_x^l. $$ 
Then we have
\bea && A|_{u=0} =
\sum_{k=1}^m( a_{k3}+c_k)\part_x^k
=\sum_{k=1}^{m-1} a_{k3}\part_x^k
+\sum_{k=1}^mc_k\part_x^k,\label{A|_{u=0}} \\ &&
B|_{u=0} =
\sum_{l=1}^n(  {b}_{l3}+d_l)\part_x^l
=\sum_{l=1}^{n-1}   {b}_{l3}\part_x^l
+\sum_{l=1}^nd_l\part_x^l,\label{B|_{u=0}} \eea
where $ a_{k3}=a_k|_{u=0}-c_k,\ 1\le k\le m,\ 
 {b}_{l3}=b_l|_{u=0}-d_l,\ 1\le l\le n.$ 
Then we can calculate that
\bea && \quad
[A,B]|_{u=0}=[A|_{u=0},B|_{u=0}]\nonumber \\&&
=[\sum _{ k = 1}^{m-1} a_{k3}\part_x^k
+\sum _{ k = 1}^mc_k\part_x^k,
\sum _{ l = 1}^{n-1} b_{l3}\part_x^l
+\sum _{ l = 1}^nd_l\part_x^l]\nonumber \\&&
=[\sum _{ k =1}^{m-1} a_{k3}\part_x^k,
\sum _{l   =1}^{n-1} b_{l3}\part_x^l]+
[\sum _{k  =1}^{m-1} a_{k3}\part_x^k,
\sum _{l =1}^nd_l\part_x^l]\nonumber \\&& \quad
+ [\sum_{k=1}^mc_k\part_x^k,
\sum_{l=1}^{n-1} b_{l3}\part_x^l].\label{[A,B]|_{u=0}}\eea
Based upon this equality, 
we obtain by Lemma \ref{lemma2} the required result.
$\vrule width 1mm height 3mm depth 0mm$

\begin{thm} \label{thm1}
Let $R, \, R_i,\ i\ge0$, be determined by (\ref{R}),(\ref{R_i}),
respectively. Then the Lax operator algebra
$R =\sum_{i=0}^\infty R_i$ forms a graded Lie algebra 
under the operation $\lbrack\!\lbrack 
\cdot,\cdot \rbrack\!\rbrack $ defined by (\ref{pd})
and thus the eigenvector field algebra
$E(R) =\sum_{i=0}^\infty E(R_i)$ forms the same graded Lie algebra
under the operation $[\cdot,\cdot ]$ defined by (\ref{Liep}),
where 
\be E(R)=\{X|L'[X]=[A,L],\ A\in R\},\ E(R_i)=\{X|L'[X]=[A,L],\ 
A\in R_i\},\ i\ge 0.\label{E(R)E(R_i)}\ee
\end{thm}
{\bf Proof:} 
We first prove the equality (\ref{R_iR_j}), that is,
\[
\lbrack\!\lbrack 
R_i,R_j
\rbrack\!\rbrack 
\sqsubseteq R_{i+j-1} ,\ R_{-1}=0,\ i,j\ge0,\]
which shows that $R =\sum_{i=0}^\infty R_i$ is a graded Lie algebra.
Let $A\in R_i,\, B\in R_j\ (i,j\ge 0)$ and $X\in E(R_i),\,Y\in E(R_j)$ be the
eigenvector fields of $A,B$, respectively, namely $[A,L]=L'[X]=X\part _x,\ 
[B,L]=L'[Y]=Y\part _x.$ Obviously we have
\[\lbrack\!\lbrack  A,B\rbrack\!\rbrack =A'[Y] - B'[X] +[ A,B ]\in R,\]
i.e. $\lbrack\!\lbrack  A,B\rbrack\!\rbrack$ possesses the form (\ref{Aform}), 
and thus we may assume that
\[\lbrack\!\lbrack  A,B\rbrack\!\rbrack =\sum_{r=1}^se_r\part _x^r=P(f_1,\cdots
,f_s). \]
We observe that 
\bea  && \lbrack\!\lbrack  A,B\rbrack\!\rbrack  |_{u=0}
=(A'[Y] - B'[X] +[ A,B ])  |_{u=0} \nonumber\\&&
=A'[Y|_{u=0} ]|_{u=0}  - B'[X|_{u=0} ]|_{u=0} +[ A,B ]|_{u=0} .\label{pd0p}
\eea
When $i+j=0$, i.e. $i=j=0$, it follows from the above equality, 
Lemma \ref{lemma3} and Lemma \ref{lemma4}
that $ \lbrack\!\lbrack A,B
\rbrack\!\rbrack|_{u=0} =0 $. Thus by Lemma \ref{lemma1},
we obtain 
$ \lbrack\!\lbrack  A,B \rbrack\!\rbrack =0$, i.e.
$ \lbrack\!\lbrack  A,B \rbrack\!\rbrack \in R_{-1}.$
Now we assume that
 $i+j\ge1$. Note that 
$A'[Y|_{u=0}]|_{u=0}$ and $B'[X|_{u=0}]|_{u=0}$ 
are linear with $Y|_{u=0}$ and $X|_{u=0}$, respectively.
It follows similarly from (\ref{pd0p}), Lemma \ref{lemma3} and Lemma
\ref{lemma4}
that the coefficients of the differential operator 
$ \lbrack\!\lbrack  A,B \rbrack\!\rbrack |_{u=0} $ are polynomials in
$x,y$ with degrees less than $i+j$ with respect to $y$.
It means by Lemma \ref{lemma1} that $e_r|_{u=0}=e_{r3}+f_r,\ 1\le r\le s$, are
polynomials in $x,y$ with degrees less than $i+j$ with respect to $y$.
On the other hand, by Lemma \ref{lemma2},
 the degree of $e_{r3}=e_r|_{u=0}-f_r$
  with respect to $y$ is less 
than the maximum degree of the $f_l$, $1\le l\le s$, with respect to $y$.
Therefore $f_r\in C_{i+j-1}[y],\ 1\le r\le s,$ which means that 
$ \lbrack\!\lbrack  A,B \rbrack\!\rbrack \in R_{i+j-1}$. 
In conclusion,
we see that the relation (\ref{R_iR_j}) holds. 

The second result of 
the theorem is obvious,
since we have (\ref{Laxoperatorpp}), i.e. 
$[\lbrack\!\lbrack  A,B \rbrack\!\rbrack,L]=
[X,Y]\part _x$ when $[A,L]=X\part _x$ and $[B,L]=Y\part _x$. 
Therefore the proof is complete.
$\vrule width 1mm height 3mm depth 0mm$

By noting (\ref{A|_{u=0}}), (\ref{B|_{u=0}}) and (\ref{[A,B]|_{u=0}}),
we may obtain from (\ref{pd0p}) that 
\be 
\lbrack\!\lbrack P(c_1,\cdots , c_m), P(d_1,\cdots , c_n)\rbrack\!\rbrack
=P(f_1,\cdots , f_{m+n-3},\frac n2 c_{my}d_n-\frac m2 c_md_{ny}),
\label{fristcoefficient}
\ee 
where $P(c_1,\cdots , c_m)=\sum_{k=1}^ma_k\part _x^k=\sum_{k=1}^m
(a_{k1}+a_{k2})\part _x^k$, the 
$a_{k1}$ and the $ a_{k2}$ being
defined by (\ref{a_krecursionrelation}).
However, it is very complicated to obtain the explicit expressions for the 
polynomials $f_i$. It follows from (\ref{fristcoefficient}) that 
\[ \left [ \sigma ^{\{m\}}(f),\sigma ^{\{n\}}(g)\right ]=\sigma ^{\{m+n-2\}}
(\frac n2 f_{y}g-\frac m2 fg_{y}),\ m,n\ge1,\]
where $\sigma^{\{0\}}(f)=0$, is a Lie product. The special case of $m=n=2$
leads to an interesting Virasoro-type Lie algebra
\[ \left [ \sigma ^{\{2\}}(f),\sigma ^{\{2\}}(g)\right ]=\sigma ^{\{2\}}
(  f_{y}g- fg_{y}).\]

Now let us choose the specific Lax operators.
If we choose these as
\bea && A_m =P(\underbrace {0,\cdots, 0}_m, 1)
=\sum_{k=1}^{m+1} a_k^{\{m\}} \part_x^k,\ m\ge 0,\\&&
B_{in} =P(\underbrace  {0,\cdots, 0}_n, y^i)=\sum_{l=1}^{n+1}
 b_{l}^{\{in\}} \part_x^l,\ i\ge1,\ n\ge 0,
\eea
then the corresponding eigenvector fields read as
\bea &&X_m=[A_m,L]\part ^{-1}_x=\sum_{k=1}^{m+1} a_k^{\{m\}}( \part_x^ku)-
(a_{1xx}^{\{m\}}+ua_{1x}^{\{m\}}+a_{1y}^{\{m\}}),\ m\ge 0,\label{X_m}\\
&& Y_{in}=[B_{in},L]\part ^{-1}_x=\sum_{l=1}^{n+1} b_l^{\{in\}} (\part_x^lu)-
(b_{1xx}^{\{in\}}+ub_{1x}^{\{in\}}+b_{1y}^{\{in\}}),\ i\ge1,\ n\ge 0.\qquad  
\eea
By Lemma \ref{lemma1}, we see that
\bea
&&R_0=\textrm{Span}\{A_m |\, m\ge 0\},\ R_i=
\textrm{Span}\{B_{jn}|\, n\ge 0,\ 0\le j\le i\},\ i\ge1;\\&&
E(R_0)=\textrm{Span}\{X_m |\, m\ge 0\},\ E(R_i)=
\textrm{Span}\{Y_{jn}|\, n\ge 0,\ 0\le j\le i\},\
i\ge1.\qquad \eea

The equations $u_t=X_m,\ m\ge 0,$ constitute 
the integrable modified KP
hierarchy. By Theorem \ref{thm1}, this  modified KP hierarchy has two
Virasoro algebras 
\be <R_0+R_1,
\lbrack\!\lbrack 
\cdot,\cdot
\rbrack\!\rbrack>\ \ 
\textrm{and}\  \ <E(R_0)+E(R_1), [ \cdot, \cdot ] >\ee
and two graded Lie  algebras
\be <R =\sum_{i=0}^\infty R_i,
\lbrack\!\lbrack 
\cdot,\cdot
\rbrack\!\rbrack>\ \ 
\textrm{and}\ \  <E(R) =\sum_{i=0}^\infty E(R_i),[\cdot,\cdot]>.
\label{gradedvfa}\ee
In particular, the equation $u_t=X_2$ is exactly  the normal modified KP
equation introduced in \cite{Konopelchenko} (see (\ref{X_2}) below)
and the space $E(R_1)$ includes all the master symmetries presented in 
\cite{Cheng}.

Through Theorem \ref{thm1}, we find at once that 
every  modified KP equation $u_t=X_i\, ( i\ge0)$, $X_i$
given by (\ref{X_m}),
possesses a hierarchy of
common time-independent symmetries $\{X_m\}_{m=0}^\infty$ and infinitely many
hierarchies of polynomial-in-time dependent symmetries 
\be \bigl\{ \sigma _i(Y _{kn}) =\sum _{j=0}^k \frac {t^j} {j!} 
(\textrm{ad}_{X_i})
^j Y_{kn}\bigr\}_{n=0}^\infty,\ k\ge1.\label{sigma_i(Y_{kn})}\ee
Further by applying the theory presented in the last section
, we obtain the following consequence.
A modified KP equation with $m$ given {\it arbitrary} functions 
$\al _i(t),\ 1\le i\le m$,
\be u_t=\al _1(t)X_{i_1}+\al _2(t)X_{i_2}+\cdots +\al _m(t)X_{i_m} \ee
has a graded symmetry algebra
\be <\sigma _{i_1i_2\cdots i_m}(E(R))=\sum_{i=0}^\infty
\sigma _{i_1i_2\cdots i_m}(E(R_i)),[\cdot,\cdot]>.\label{gradedsa}
\ee
The map $\sigma _{i_1i_2\cdots i_m}$ is defined by 
\bea &&
 \sigma _{i_1i_2\cdots i_m}(\rho_i)=
\textrm{exp}(\textrm{ad}_{X_{i_1}})\cdots \textrm{exp}(\textrm{ad}_{X_{i_m}})
\rho_i\nonumber \\ &&
=\sum_{0\le j_1+j_2+\cdots +j_m\le i}
\frac {\beta _1^{j_1}\beta _2^{j_2}\cdots \beta _m^{j_m}}{j_1!j_2!\cdots j_m!}
(\textrm{ad}_{X_{i_1}})^{j_1}(\textrm{ad}_{X_{i_2}})^{j_2}\cdots 
(\textrm{ad}_{X_{i_m}})^{j_m} \rho_i,\ \rho_i\in E(R_i), \qquad \eea 
where $\frac {\part }{\part t}\beta _j(t)=\al _j(t),\ 1\le j\le m$.
Moreover this map is a Lie algebra homomorphism between the graded 
vector field algebra defined in (\ref{gradedvfa}) and the graded symmetry 
algebra defined by (\ref{gradedsa}). In other words, we have
\be \left\{ \ba {l} 
 \left[\sigma _{i_1i_2\cdots i_m}(X_{m_1}),\sigma _{i_1i_2\cdots i_m}
(X_{m_2})\right]=0,\vspace{2mm} \\ 
\left[\sigma _{i_1i_2\cdots i_m}(X_{m_1}),\sigma _{i_1i_2\cdots i_m}
(Y_{j_1n_1})\right]=\sigma _{i_1i_2\cdots i_m}
(\left[X_{m_1},Y_{j_1n_1}\right]),\vspace{2mm}
 \\ 
\left[\sigma _{i_1i_2\cdots i_m}(Y_{j_1n_1}),\sigma _{i_1i_2\cdots i_m}
(Y_{j_2n_2})\right]=\sigma _{i_1i_2\cdots i_m}
(\left[Y_{j_1n_1},Y_{j_2n_2}\right]).
\ea \right.
\ee
The symmetries $\sigma _{i_1i_2\cdots i_m}(Y_{in})$ contain $m$ given 
arbitrary functions of time and thus the polynomial-in-time symmetries 
(\ref{sigma_i(Y_{kn})})
generated by the master symmetries $Y_{kn}$ 
are no more than special cases amongst these.  

Some concrete examples of the Lax operators $A$, $B$ and the corresponding 
eigenvector fields $X$, $Y$ are listed in the following:- 

\noindent {(i) The Lax operators and the vector fields of the modified KP 
equations:}
\bea  A_0&=&P(1)=\part _x,\ X_0=u_x;
\nonumber \\ 
A_1&=&P(0,1)=\part _x^2+u\part _x,\ X_1=-u_{y};\nonumber \\ 
A_2&=&P(0,0,1)=\sum_{k=1}^3a_{k}^{\{2\}}\part _x^k=\part _x^3+
\frac 32 u\part _x^2 + (\frac 38 u^2+\frac 34 u_x -\frac 34 \part ^{-1}_x
u_y)\part _x, \nonumber \\ 
X_2&= &\sum_{k=1}^3a^{\{2\}}_k(\part _x^ku)-(a_{1xx}^{\{2\}}+ua_{1x}^{\{2\}}+
a_{1y}^{\{2\}})\nonumber  \\ 
&=&\frac 14 u_{xxx}-\frac 38 u^2u_x -\frac 34 u_x\part ^{-1}_x u_y 
+\frac 34 \part ^{-1}_x u_{yy}. \label{X_2}
\eea
\noindent {(ii) The Lax operators and the $i$-th master symmetries 
of the modified KP equations:}
\bea 
B_{i0}&=&P(y^i)=y^i\part _x=y^iA_0 ,\ Y_{i0}=y^iu_x-iy^{i-1}=y^iX_0-iy^{i-1};
 \nonumber \\ 
B_{i1}&=&\sum_{l=1}^2b_l^{\{i1\}}\part _x^l=y^i\part ^2_x+(-\frac12 
ixy^{i-1}+y^iu)\part _x=
y^iA_1-\frac 12 ixy^{i-1}A_0,
\nonumber \\
Y_{i1}&=&\sum_{l=1}^2b_l^{\{i1\}}
(\part _x^lu)-(b_{1xx}^{\{i1\}}+ub_{1x}^{\{i1\}}
+b_{1y}^{\{i1\}})
\nonumber \\ 
&=&-y^iu_y-\frac12 ixy^{i-1}u_x-\frac12 iy^{i-1}u+\frac12 i(i-1)xy^{i-2}
\nonumber \\ &&
=y^iX_1-\frac 12 ixy^{i-1}X_0-\frac12 iy^{i-1}u+\frac12 i(i-1)xy^{i-2}.
\nonumber
\eea  
From these, 
we see that $u_t=X_2$ is, indeed, the normal modified KP equation introduced 
in \cite{Konopelchenko}.
It is also a two-dimensional generalization of the modified KdV equation
and may be connected with the KP equation by a Miura
transformation \cite{Konopelchenko}.

\section{Concluding remarks}
\setcounter{equation}{0}

A main result is the construction of the graded symmetry algebras
for a broad class of
 variable-coefficient evolution equations with given arbitrary 
time-dependent functions as coefficients, starting from known
graded Lie algebras, in particular, centreless Virasoro algebras.
A Lie homomorphism 
$\textrm{exp}(\textrm{ad}_T)$ of the vector field Lie algebra
plays a central role in the construction 
of time-dependent symmetries of these equations. 
On the other hand,
a graded Lax operator algebra and a graded
symmetry algebra are presented for the modified KP hierarchy, 
in the style of the Lax operator method of
\cite{Ma1992a}. This gives a concrete example of graded Lie algebras.
Furthermore the time-dependent symmetries are 
obtained for the variable-coefficient modified KP equations as an application
of our main result.

We point out that coupled systems of
soliton equations may be constructed by perturbation
\cite{MaF/PhysLettA} \cite{MaF/ChaosFS} 
and thus various new realization of graded symmetry algebras
may be presented for integrable coupled systems, based upon the theory
presented in this paper. Moreover by applying the perturbation iteratively,
we may obtain infinitely many new realizations of graded Lie algebras, 
by starting from a known one.

There are also some other symmetries which may be constructed. From 
the scale transformations, for example,
we may take $T_i=x_iu_{x_i}, \ x=(x_i),$ as a key vector field.
The application of $\textrm{exp}(\textrm{ad}_{T_i})$ to the known particular 
equations may lead to some important equations with specific space-dependent 
coefficients.
An interesting remaining problem is whether or not
we can construct integrable evolution equations with {\it given arbitrary 
space-dependent functions} as coefficients and in what fashion we can 
construct their symmetries with space-dependent functions as coefficients if 
they exist.
Of course, we may also ask whether or not there is any important application
of these various equations with variable-coefficients to physical 
problems - such as,
for example, to applicable conformal field theory. All of these ideas need 
further investigation.

\vskip 2mm

\noindent{\bf Acknowledgments:} 
One of the authors (WXM) 
would like very much to thank the 
Alexander von Humboldt Foundation for 
the financial support, which 
made his visit to UMIST possible.
He is also greatly indebted to Prof. B. Fuchssteiner and Dr. W. Oevel
for kind and stimulating discussions.


\small 

\begin{thebibliography}{99}
\bibitem{OevelF} Oevel W. and Fokas A. S., 
Infinitely many commuting symmetries
and constants of motion in involution for explicitly time-dependent
evolution equations, {\it J. Math. Phys.}, 1984, Vol.25, No.4, 918--922.
\bibitem{Fuchssteiner1983} Fuchssteiner B., 
Mastersymmetries, higher order time-dependent symmetries 
and conserved densities of nonlinear evolution equations,
{\it Prog. Theor. Phys.}  1983, Vol.70, 1508--1522  
\bibitem{Fokas} Fokas A. S.,
Symmetries and integrability,
{\it Studies in Appl. Math.}, 1987, Vol.77, 253--299.
\bibitem{Ma1991}W. X. Ma, Generators of vector fields and  
time dependent symmetries of evolution equations, {\it Science in China A},
1991, Vol.34, No.7, 769--782.
\bibitem{FuchssteinerO}  Fuchssteiner B. and Oevel W., 
New hierarchy of nonlinear completely integrable systems 
related to a change of variables for evolution parameters, 
{\it Physica A}, 1987, Vol.145, 67--95.
\bibitem{Fuchssteiner1993} Fuchssteiner B., Integrable nonlinear
evolution equations with time-dependent coefficients, {\it J. Math. 
Phys.}, 1993, Vol.34, No.11, 5140--5158.
\bibitem{Ma1992a} Ma W. X.,
The algebraic structures of isospectral Lax
operators and applications to integrable equations, 
{\it J. Phys. A: Math. Gen.}, 1992, Vol.25, 
5329--5343. 
\bibitem{Fuchssteiner1996} Fuchssteiner B., Compatibility in abstract 
algebraic structures, in: Algebraic Aspects of Integrable Systems:
In Memory of Irene Dorfman, eds. Fokas A. S. and Gelfand I. M.,
Boston, Birk\"auser, 1996, p131--141.
\bibitem{Fushchych} Fushchich (Fushchych) W. I., Shtelen W. M. and 
Serov N. I., Symmetry Analysis and Exact 
Solutions of Equations of Nonlinear Mathematical Physics, Dordrecht, Kluwer,
1993.
\bibitem{FushchychT} Fushchych W. I. and Tsyfra I., On new Galilei- 
and Poincare-invariant nonlinear equations for electromagnetic field,
{\it J. Nonlinear Math. Phys.}, 1997, Vol.4, Nos.1-2, 44--48.
\bibitem{MaBCF} Ma W. X., Bullough R. K., Caudrey P. J. and Fushchych W. I.,
Time-dependent symmetries of variable-coefficient
evolution equations and graded Lie algebras. {\it To be published}.
\bibitem{ChenL} Chen H. H. and Lee Y. C., 
A new hierarchy of symmetries for the integrable
nonlinear evolution equations,
in: Advances in Nonlinear Waves Vol. II, Research 
Notes in Mathematics 111, ed. L. Debnath, Boston, Pitman, 1985, p233-239.
\bibitem{MaF} Ma W. X. and Fuchssteiner B., 
Algebraic structure of 
discrete zero curvature equations and 
master symmetries of discrete
evolution equations, 1996, preprint.
\bibitem{ChengLB} Cheng Y., Li Y. S. and Bullough R. K.,
Integrable non-isospectral flows associated with the Kadomtsev-Petviashvili
equations in $2+1$ dimensions,
{\it J. Phys. A: Math. Gen.}, 1988, Vol.21, L443-L449.
\bibitem{Ma1990} Ma W. X., $K$-symmetries and $\tau $-symmetries of
evolution equations and their Lie algebras, 
{\it J. Phys. A: Math. Gen.}, 1990, Vol.23, 2707--2716.
\bibitem{OevelFZ} Oevel W., Fuchssteiner B. and Zhang H. W.,  
Mastersymmetries and mutil-Hamiltonian formulations for some integrable
lattice systems,
{\it Prog. Theor. Phys.}, 1989, Vol.81, 294--308.
\bibitem{MaZ} Ma W. X. and Zhou Z. X., Coupled integrable systems 
  associated with a polynomial 
  spectral problem and their Virasoro symmetry algebras, {\it 
Prog. Theor. Phys.}, 1996, Vol.96, No.2, 449--457.
\bibitem{Ma1992} Ma W. X., 
Lax representations and Lax operator algebras 
of isospectral and nonisospectral hierarchies of evolution equations, 
{\it J. Math. Phys.}, 1992, Vol.33, 2464--2476.
\bibitem{Kac} Kac V. G., Infinite Dimensional Lie Algebras 3rd ed.,   
Cambridge, Cambridge University Press, 1990.
\bibitem{BulloughC1} Bullough R. K. and Caudrey P. J., 
Solitons and the Korteweg-de Vries equation: 
Integrable systems in 1834-1995, {\it Acta Appl. Math.}, 1995, Vol.39, 
193--228 and references.
\bibitem{BulloughC2} Bullough R. K. and Caudrey P. J., 
Symmetries of the classical integrable systems and 2-dimensional quantum
gravity: a `map', {\it J. Nonlinear Math. Phys.}, 1996, Vol.3, 245--259. 
and references.
\bibitem{ChenLL} Chen H. H., Lee Y. C. and Lin J. E., 
On a new hierarchy of symmetries for the Kadomtsev-Petviashvili equation,
{\it Physica D}, 1983, Vol.9, 439--445.
\bibitem{Konopelchenko} Konopelchenko B. G., On the gauge-invariant description
of the evolution equations integrable by Gelfand-Dikij spectral problems,
{\it Phys. Lett. A}, 1982, Vol.92, No.7, 323--327.
\bibitem{Cheng} Cheng Y., Hierarchies of nonlinear integrable equations
and their symmetries in $2+1$ dimensions, {\it Physica D}, 1990,
Vol.46, 286--294. 
\bibitem{MaF/PhysLettA} Ma W. X. and Fuchssteiner B., 
The bi-Hamiltonian structures 
of the perturbation equations of KdV hierarchy, {\it Phys. Lett. A}, 1996,
Vol.213, 49--55.
\bibitem{MaF/ChaosFS} Ma W. X. and Fuchssteiner B., Integrable theory 
of the perturbation equations, {\it  Chaos, Solitons and Fractals}, 1996, 
Vol.7, No.8, 1227--1250.
\end{thebibliography}
\end{document}